\documentclass[prb,superscriptaddress,twocolumn,floats,floatfix,]{revtex4}
\usepackage{graphicx}
\usepackage{mciteplus}
\usepackage{gensymb}
\usepackage{amsmath}
\usepackage{chemformula} 
\usepackage[T1]{fontenc} 
\usepackage{upgreek}
\usepackage{color}

\begin{document}


\title{Gigahertz Single-Electron Pumping Mediated by Parasitic States}

\author{A. Rossi}
\email[Electronic mail: ]{ar446@cam.ac.uk}
\affiliation{Cavendish Laboratory, University of Cambridge, J.J. Thomson Avenue, Cambridge, CB3 0HE, U.K.}
\author{J. Klochan}
\author{J. Timoshenko}
\affiliation{Faculty of Physics and Mathematics, University of Latvia, Riga LV-1002, Latvia}
\author{F. E. Hudson}
\affiliation{School of Electrical Engineering \& Telecommunications, The University of New South Wales, Sydney 2052, Australia}
\author{M. M\"{o}tt\"{o}nen}
\affiliation{QCD Labs, QTF Centre of Excellence, Department of Applied Physics, Aalto University, PO Box 13500, FI-00076 Aalto, Finland}
\author{S. Rogge}
\affiliation{School of Physics, The University of New South Wales, Sydney 2052, Australia}
\author{A. S. Dzurak}
\affiliation{School of Electrical Engineering \& Telecommunications, The University of New South Wales, Sydney 2052, Australia}
\author{V. Kashcheyevs}
\affiliation{Faculty of Physics and Mathematics, University of Latvia, Riga LV-1002, Latvia}
\author{G. C. Tettamanzi}
\email[Electronic mail: ]{giuseppe.tettamanzi@adelaide.edu.au}
\affiliation{Institute of Photonics and Advanced Sensing and School of Physical Sciences, The University of Adelaide, Adelaide SA 5005, Australia}
\date{\today}

\begin{abstract}
In quantum metrology, semiconductor single-electron pumps are used to generate accurate electric currents with the ultimate goal of implementing the emerging quantum standard of the ampere. Pumps based on electrostatically defined tunable quantum dots (QDs) have thus far shown the most promising performance in combining fast and accurate charge transfer. However, at frequencies exceeding approximately 1 GHz, the accuracy typically decreases. Recently, hybrid pumps based on QDs coupled to trap states have led to increased transfer rates due to tighter electrostatic confinement. Here, we operate a  hybrid electron pump in silicon obtained by coupling a QD to multiple parasitic states, and achieve robust current quantization up to a few gigahertz. We show that the fidelity of the electron capture depends on the sequence in which the parasitic states become available for loading, resulting in distinctive frequency dependent features in the pumped current.
\end{abstract}

\maketitle
Semiconductor quantum dots (QDs) with tunable tunnel barriers~\cite{kou,wiel,hanson_rev,floris} are routinely used to realise single-electron pumps.\cite{slava-rev} These devices have potential applications as single-particle emitters for fermionic optics,\cite{bocq,jon,niels} as well as sources of quantized electric current.\cite{pek-rev} The use of QDs for the generation of highly accurate current has arguably become the most promising direct route for the practical realization of the new quantum ampere.\cite{mills} Indeed, in the last few years, QD pumps in both GaAs and Si systems have been operated at increasing driving frequencies and lower than part-per-million (ppm) uncertainty,\cite{stein-apl,stein-met,gento-apl,steve-met,zhao} rapidly approaching the stringent metrological requirements.\cite{hans} \\\indent
Alongside this generation of pumps based on electrostatically defined QDs, devices based on atomic impurities in silicon have also emerged and have demonstrated current quantization capabilities.\cite{lans,roche,gct,gct-sr}  The operation of these devices may be less challenging than that of the  QD-based pumps, because of high operation temperatures, simplified device layout and no need for external electric or magnetic confinement. However, it is not yet clear whether the necessary transfer speed and accuracy can be attained, given the intrinsically limited tunability of these systems. Recently, hybrid pumps, where unintentional trap states contribute to the electron transfer together with an electrostatically defined QD, have been reported.\cite{gento-ncomm,wenz,gento-sr,clapera} Some of these systems have demonstrated very high speed of operation~\cite{gento-ncomm} and a certified accuracy at the level of a few tens of ppm.\cite{gento-sr} The tunability of these devices casts a positive outlook on the possibility of improving the accuracy further, while keeping the operation speed in the gigahertz range. Furthermore, the study of these systems may shed light on some of the yet unexplained mechanisms responsible for the occurrence of pumping errors at high frequencies.\cite{gento-sr,gct-sr}\\\indent
Here, we present an alternative realization of a hybrid pump, where parasitic states localised at a tunnel barrier compete with the intended QD pump during the electron loading and capture phase of the pumping cycle. Our system, fabricated using highly tunable silicon metal-oxide-semiconductor (MOS) technology,\cite{jove} allows one to adjust the number of captured and ejected electrons per cycle. Consequently, we manage to achieve accurate current quantization up to a few gigahertz of operational frequency. Importantly, an analytical model is developed and used to elucidate the competition of the parasitic states in the capture dynamics. Our findings indicate that the capture fidelity depends on the sequence in which the states become available, leading to distinctive frequency-dependent features in the pumped current. Finally, we experimentally test the flatness of the current plateau at 2.5 GHz and observe quantization within the sub-ppm statistical uncertainty.
\begin{figure}[t]
\includegraphics[scale=0.8]{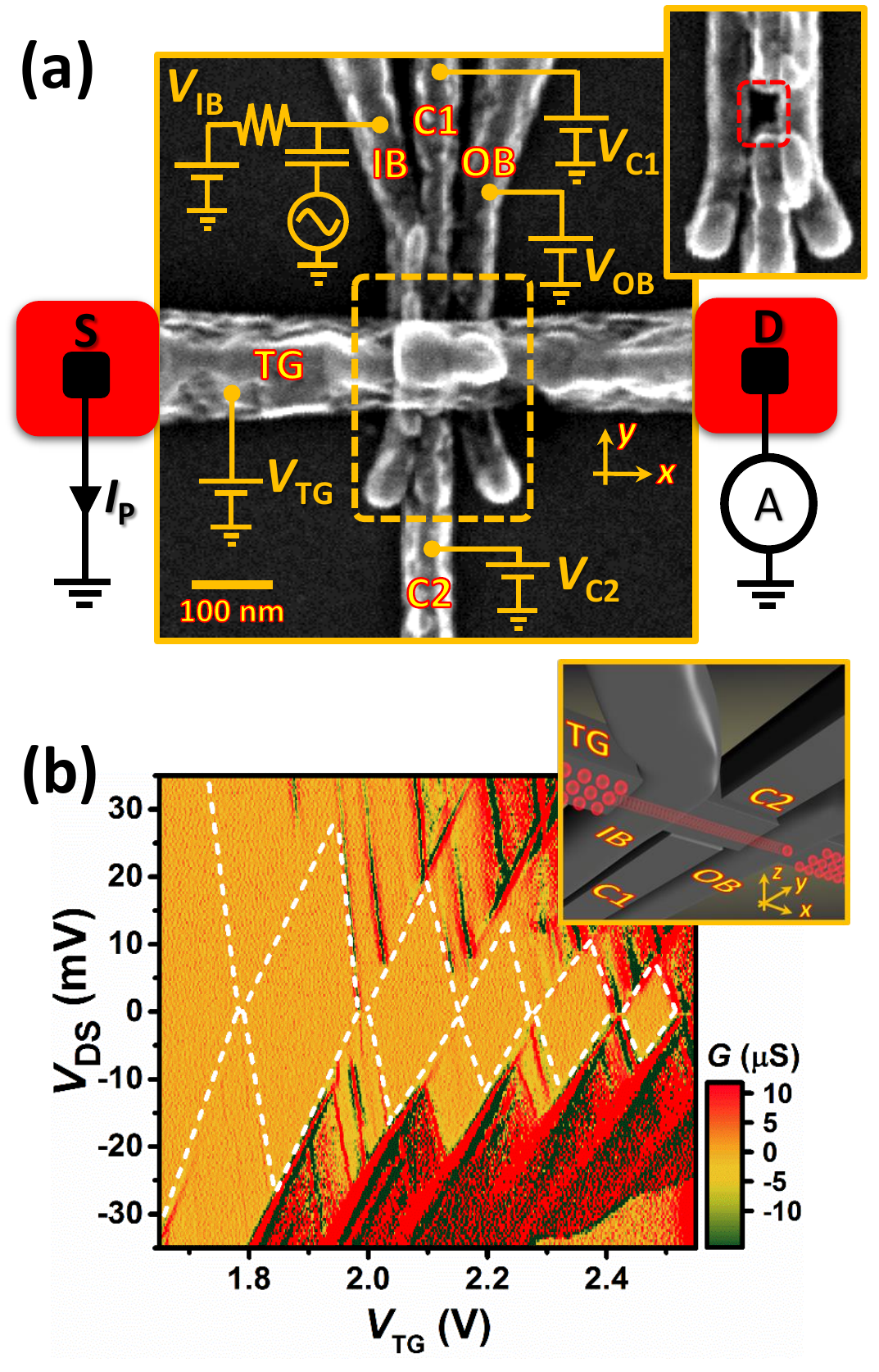}
\caption{(a) Scanning electron microscope (SEM) image of a device similar to that used in the experiments. The red pads represent the source (S) and drain (D) ohmics. The measurement set-up is also sketched. Inset: SEM image of the area enclosed by the yellow dashed line in the main panel. The picture is taken before the third aluminum layer is deposited, and allows one to observe the central region where the quantum dot is formed (region bounded by the red dashed line). The lengthscale is the same as in the main panel.  (b) Differential conductance of the device as a function of top gate and source--drain voltages. Dashed lines are guides for the eye to highlight the boundaries of the Coulomb diamonds. The other voltages are given by: $V_\textup{IB}=0.78$~V, $V_\textup{OB}=0.95$~V, $V_\textup{C1}=0.15$~V, and $V_\textup{C2}=-0.49$~V. Inset: Schematic illustration of the device architecture with gate electrodes in grey and electrons in red.}
\label{set-up}
\end{figure}

The device used for this study has been fabricated by means of multi-gate MOS technology on silicon.~\cite{jove} The high-purity, near-intrinsic  silicon substrate has n+ ohmic regions for source and drain contacts defined by phosphorous diffusion. The gate stack is made of three layers of Al/Al$_y$O$_x$ electrodes patterned using electron-beam lithography, Al thermal evaporation and oxidation. These are deposited on a high-quality 5-nm-thick SiO$_2$ gate oxide, which is thermally grown in a dry atmosphere at 800~$^{\circ}$C. With respect to previous single-electron pump realizations in planar MOS technology,~\cite{mynano,zhao,scirep} we have improved upon the gate design, in order to decrease the total number of gates needed per device, as well as to reduce the QD size. As illustrated in Fig.~\ref{set-up}(a), this device has five gates, as opposed to previous reports where seven electrodes were necessary to control the QD occupancy and tune the lateral confinement.~\cite{mynano,zhao} Furthermore, by careful alignment of the first and second metal depositions, an area as small as 30$\times$30 nm$^2$ defines the QD, as shown in the inset of Fig.~\ref{set-up}(a). This is achieved by using gate IB (OB) to form a tunnel barrier at the entrance (exit) of the pump, and by creating a strong lateral confinement with gates C1 and C2. The third layer defines a top gate (TG), which controls both the density of states in the leads and the potential of the dot. In order to reach the few-electron regime without depleting the reservoirs, we use one of the confinement gates to control the occupancy of the dot, whilst $V_\textup{TG}$ is set well above the threshold voltage for conduction in the leads. By operating the sample at cryogenic temperatures and with appropriate tuning of the gate voltages, we observe single-electron tunnelling and measure a dot charging energy in excess of 25 meV, as shown in Fig.~\ref{set-up}(b). This is roughly a 50\% increase with respect to the most compact QD pumps~\cite{gento-apl,zhao}, and it is notably attained with reduced device complexity. A tight charge confinement is desirable because it results in good energy separation between charge states and, hence, improved resilience against pumping errors.~\cite{mynano,kriss}\\\indent
In an effort to lift experimental challenges related to single-electron pumping, the device was operated at a conveniently accessible 4.2-K temperature of a liquid helium cryostat. Flexible coaxial cables fitted with low-temperature low-pass filters were employed to connect the device with the room-temperature electronics. In order to reduce pick-up noise, the gates were biased by battery-powered voltage sources. The pumping experiments were carried out by connecting gate IB to a low-temperature RC bias tee, which allows one to superimpose an ac signal to a dc bias. The driving gate was connected to a high-frequency semirigid coaxial cable attenuated by 6 dB at low temperature. A single sinusoidal drive in the absence of both source-drain bias and magnetic field was used for pumping. The generated electric current was measured with a transimpedance amplifier at room temperature.

\begin{figure}[]
\includegraphics[scale=0.75]{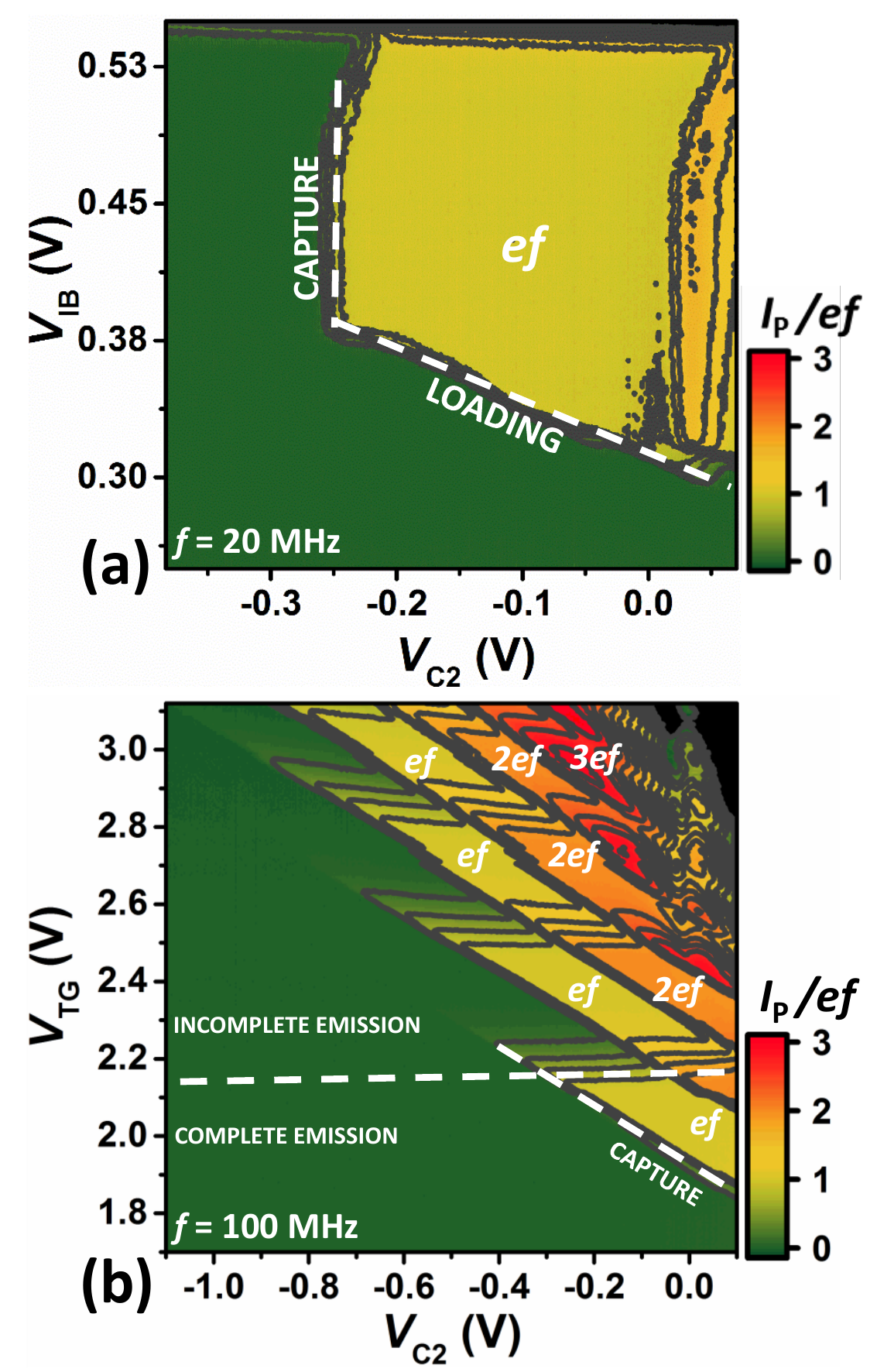}
\caption{(a) Pumped current as a function of the entrance barrier and $\textup{C2}$ gate voltages at $f=20$~MHz. White dashed lines are guides for the eye to highlight the electron loading and capture boundaries. The other experimental parameters: $V_\textup{C1}=0.15$~V, $V_\textup{OB}=1.05$~V, $V_\textup{TG}=1.9$~V, and $P_\textup{ac}=0.3$~dBm. (b) Pumped current as a function of top gate and $\textup{C2}$ gate voltages at $f=100$~MHz. White dashed lines are guides for the eye to highlight the emission and capture boundaries. The other experimental parameters: $V_\textup{C1}=0.1$~V, $V_\textup{OB}=1.03$~V, $V_\textup{IB}=0.46$~V, and $P_\textup{ac}=0$~dBm.}
\label{check}
\end{figure}

\begin{figure*}[]
\includegraphics[scale=0.65]{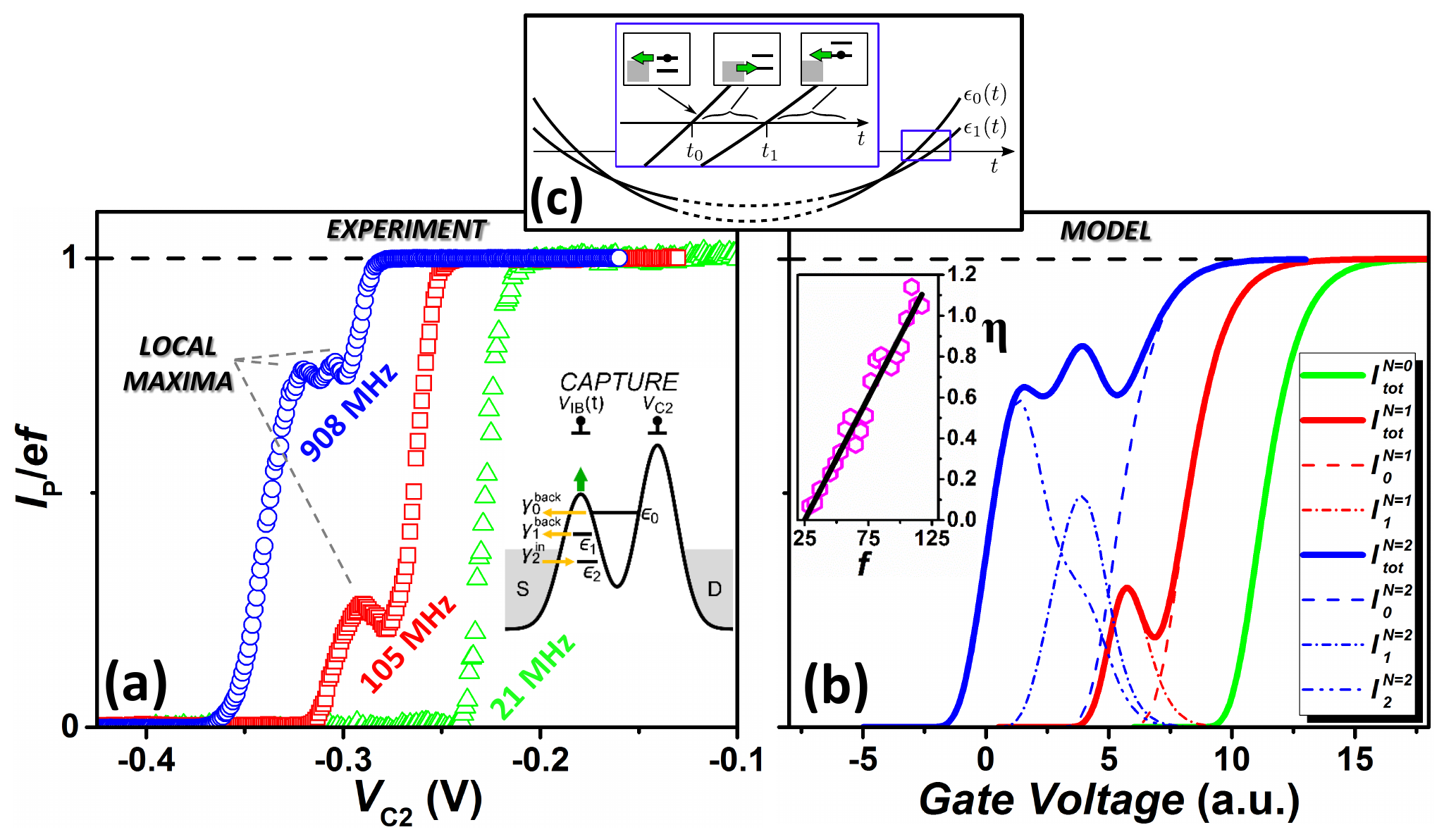}
\caption{(a) Measured pumped current as a function of gate $\textup{C2}$ voltage for three different driving frequencies. The curves are shifted horizontally for clarity. The dashed line is a guide for the eye to indicate the expected quantized current values. Experimental parameters at $f=21$~MHz and $f=105$~MHz: $V_\textup{C1}=0.1$~V, $V_\textup{OB}=1.2$~V, $V_\textup{IB}=0.52$~V, $V_\textup{TG}=2.0$~V, and $P_\textup{ac}=0.3$~dBm. Experimental parameters at $f=908$~MHz: $V_\textup{C1}=0.2$~V, $V_\textup{OB}=0.98$~V, $V_\textup{IB}=0.51$~V, $V_\textup{TG}=1.9$~V, and $P_\textup{ac}=3.6$~dBm. Inset: Potential landscape illustrating three states involved during the initialisation phase. In-tunnelling and back-tunnelling rates from/to the source electrode and the energy levels are sketched. (b) Calculated current values as functions of the control voltage $v$ given by the model. Model parameters: $N=0$ (green), $N=1$ (red, $\delta = 2.4$, $\eta=2.2$) and  $N=2$ (blue, $\delta = 2.6$, $\eta=18$). Dotted lines show 
individual state contributions. Inset: The parameter $\eta$ extracted from the experimentally measured values of the height of the first local maximum for $29 \leq f \leq 117$~MHz using Eq.~\eqref{eq:Ipeak2}. The solid line is a fit to $\eta \propto (f-f_\textup{c})$ which gives $f_\textup{c} =24.9 $ MHz. (c) Schematic diagram of the time evolution of the single-electron energy levels for the main ($\epsilon_0$) and the parasitic ($\epsilon_1$) dot.
The horizontal axis is drawn at the location of the source Fermi level. The pictograms on an enlarged timescale illustrate the processes responsible for loading-limited current through the parasitic dot, as modelled in Eq.~(1).}
\label{cmp}
\end{figure*}

By driving the pump at a relatively low frequency, $f$, and sufficiently large power, $P_\textup{ac}$, one can identify regions of the gate voltage parameter space where robust current quantization occurs. Figure~\ref{check}(a) shows the pumped current normalized to the expected quantized value, $\mathrm{e} f$, as a function of the voltages applied to the entrance barrier gate and to one of the confinement gates. The contour lines reveal a large region of quantized current. Its boundaries for insufficient loading and capture of the first electron are compatible with the typical characteristics of tunable-barrier QD pumps.\cite{slava-rev} Note that, in our particularly compact device design, both confinement gates affect the transparency of the tunnel barriers.
In the topmost region of Fig.~\ref{check}(a) one typically expects to observe a current reduction with a boundary determined by insufficient electron emission.\cite{slava-rev} In contrast, we measure a large current that we ascribe to rectification effects arising from the fact that the large-amplitude ac drive takes the instantaneous operation point well into the region of full turn-on for the entrance barrier.\cite{spg-jap} In order to optimise the robustness of the pump, in the following we will choose $V_\textup{IB}$ such that the operation point is in the vicinity of neither the limit for insufficient loading nor the current rectification.\\\indent
Figure~\ref{check}(b) shows the pumped current as a function of $V_\textup{TG}$ and $V_\textup{C2}$. In this case, we observe incomplete emission without large rectified currents. The capture line shows a negative slope, as expected from the fact that both gates efficiently control the QD potential. However, whereas TG has virtually no effect on the tunnel barriers due to screening from the metal layers below, $V_\textup{C2}$ influences the exit barrier potential. Hence, for increasingly negative $V_\textup{C2}$, the output barrier potential grows sufficiently to prevent charge emission at a fixed drive amplitude. Emission, albeit incomplete, can be restored by increasing the QD occupancy for large $V_\textup{TG}$. The resulting diagram in Fig.~\ref{check}(b) is typical for tunable-barrier pumps in which the number of electrons captured and ejected per cycle can be independently adjusted.\cite{fuji04,fuji08} Our pump shows robust quantization and tunability of the generated current up to 3.55 GHz (see Supporting Information).\\\indent
We tune the operation of the pump in the complete emission regime and turn to discuss the capture characteristics as a function of the driving frequency. Figure~\ref{cmp}(a) shows the pumped current as a function of $V_\textup{C2}$ for three different driving frequencies. 
For $f=21$~MHz, the current raises monotonically from $I_\textup{P}=0$ to $I_\textup{P}=\mathrm{e} f$, as one would expect from a capture process of increasing fidelity. This is consistent with a decay cascade initialization with back-tunnelling rates exponentially dependent on $V_\textup{C2}$.~\cite{slava-prl} At higher frequencies, the rising edge of the current characteristics becomes non-monotonic. In particular, we observe one local maximum at $f=105$~MHz and two local maxima at $f=908$~MHz. Data for a wider range of driving frequencies are included in the Supporting Information. In the previous literature,\cite{mynano,kata} the effect of increasingly high frequency of the drive resulted in the appearance of additional plateaus on the rising edge of the main current plateau. This was attributed to non-adiabatic excitations of the captured electron to high-energy states in the QD spectrum. We stress that this effect cannot explain our findings, particularly the observation of local current maxima as opposed to additional plateaus. Hence, it is clear that in our system a new and yet to be explained phenomenon is in place. To understand the underlying physical mechanisms, we have developed the following interpretation.\\
 \begin{figure*}[t]
\includegraphics[scale=0.6]{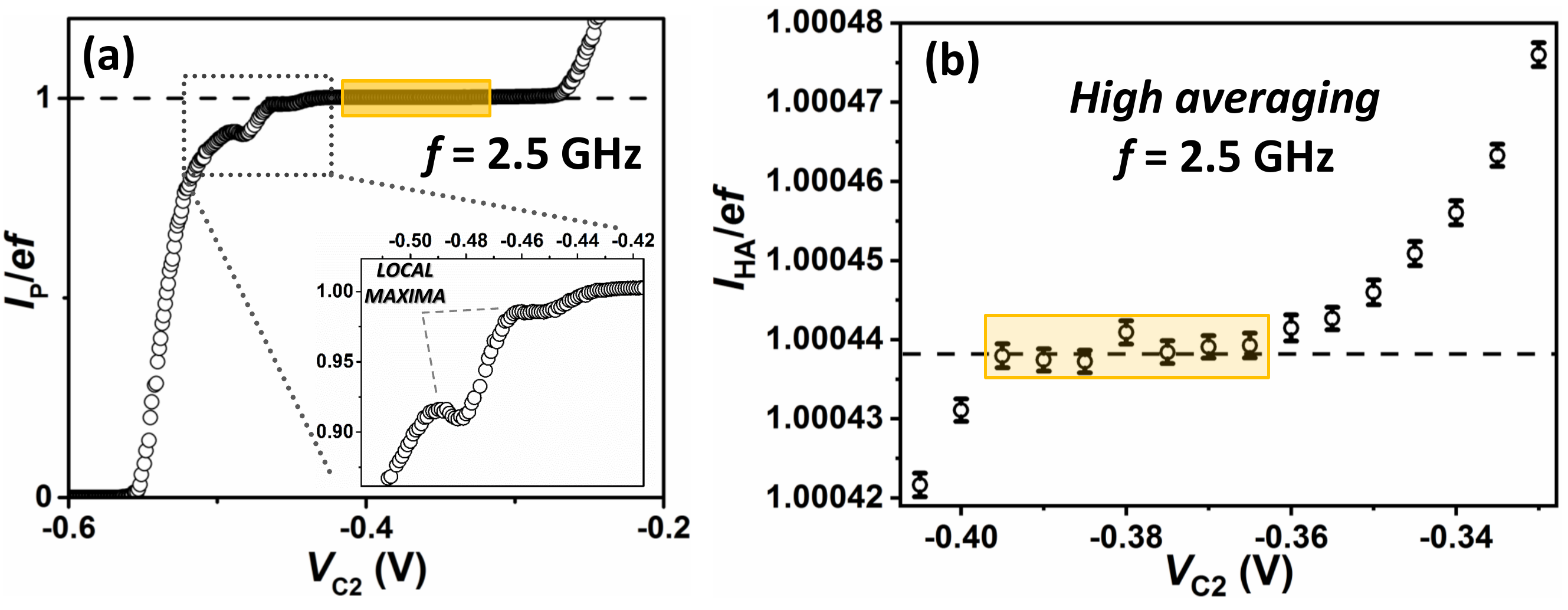}
\caption{(a) Pumped current as a function of $\textup{C2}$ gate voltage. The dashed line is a guide for the eye to indicate the expected quantized current value. The area in yellow represents the voltage range where the high-averaging measurements of panel (b) are carried out. Experimental conditions: $f=2.5$~GHz, $P_\textup{ac}=3.7$~dBm, $V_\textup{IB}=0.49$~V, $V_\textup{OB}=1.08$~V, $V_\textup{TG}=1.99$~V, and $V_\textup{C1}=-0.28$~V. Inset: a selected subset of the data from the main panel in an enlarged scale. Local maxima in the rising edge of the current plateau are clearly visible. (b) High-averaging measurement of the normalized pumped current as a function of $V_\textup{C2}$. Circles (error bars) represent the mean ($1\sigma$ random uncertainty) of  readings taken over 12 on/off cycles. The dashed line indicates the mean of the distribution of the 7 points on the plateau enclosed by the yellow shaded area. The experimental conditions are identical to those in panel (a).}
\label{accu}
\end{figure*}
Let us first briefly summarise the conventional model of back-tunnelling-dominated capture for a single-QD pump.~\cite{slava-prl,Fricke13,slava-rev} The dot is assumed to be well-defined and filled with one electron before time moment $t_0$ when
the chemical potential of this singly occupied dot crosses the Fermi level of the source lead on the way up, thus triggering the back-tunnelling.
Provided that the emission into the drain is complete, the pumped current is then dominated by the capture probability, 
$I_\textup{P}/(\mathrm{e}f) = P_0^{\textup{C}} = \exp (-X_0)$ where 
$X_0 = \int_{t_0}^{t^{\ast}} \gamma_0^{\text{back}}(t) \,  d t $ is the integrated back-tunnelling rate. A time $t^{\ast}>t_0$  within the pumping cycle separates the
capture from the subsequent emission, with the condition that  $\gamma_0^{\text{back}}(t)$ is negligible for $t>t_{\ast}$. 
More positive $V_\textup{C2}$ results in tighter confinement (i.e. smaller $\gamma_0^{\text{back}}(t_0)$), and, hence,  $X_0$ 
is expected to decrease with increasing $V_\textup{C2}$. Typically, $X_0 \propto e^{-v}$ where the dimensionless voltage parameter $v$ is a linear function of $V_\textup{C2}$, resulting in a characteristic double-exponential shape of current quantization steps,~\cite{slava-prl} as we observe for $f=21$~MHz.\\\indent
We argue that the features observed in the current traces of Fig.~\ref{cmp}(a) can be attributed to parasitic localized states that have lower single-electron energies and more pronounced confinement than the intended QD. Strong intra- and inter-dot Coulomb repulsion limits the total occupancy to unity and forces multiple pumps driven in parallel to compete for the capture of the single electron as the system is tuned to the vicinity of $0$-to-$1 \, \mathrm{e} f$ transition. 
Tuning $V_{C2}$ to more positive values lowers the localized energy levels faster than it decreases the height of the tunnelling barrier, hence both the back-tunnelling (pump to source) and in-tunnelling (source to pump) rates are reduced. However, the expected effect on the pumping current is opposite: the current increases ($\partial I_\textup{P}/\partial V_{\textup{C2}} >0$) for back-tunnelling-limited capture fidelity, and decreases ($\partial I_\textup{P}/\partial V_{\textup{C2}} <0$) if the probability to pump an electron is limited by loading. To illustrate the proposed mechanism for the anomalous decrease in current with increasing confinement, let us introduce the presence of one parasitic state ($N=1$) with back-tunnelling rate $\gamma_1^{\text{back}}(t) \ll \gamma_0^{\text{back}}(t)$ for $t>t_1$ and an in-tunnelling rate $\gamma_1^{\text{in}}(t)$ for $t_0<t<t_1$. Here $t_1 = t_0 + \Delta t$ is  the time when the energy level of state $1$ crosses the Fermi level. If the electron is initially in state $1$, then it is captured with probability $e^{-X_1}=\exp [-\int_{t_1}^{t^{\ast}} \gamma_1^{\text{back}}(t) \,  d t ]$ as described above, and contributes to the pumping current. However, if the electron is initially in the main dot (state $0$),  it back-tunnels immediately after $t_0$ and
makes state 1 available for loading. Here, we consider a range of $v$ such that the main dot $0$ is too open to keep the electron above the Fermi sea, i.e. $X_0 \gg 1$, but $ X_1 \sim 1$. The probability of loading into the empty parasitic state during the time from $t_0$ to $t_1$ is then simply
$1-\exp[-\int_{t_0}^{t_1} \gamma_1^{\text{in}}(t) \, dt ]$.
Adding the probabilities of the two alternative ways  to trap the electron in state 1 gives an approximation for the total pumped current
\begin{align} \label{eq:eq1}
 \frac{I_P}{\mathrm{e}f}&=\left [ P_{1}^{L}+ P_{0}^{L} \left(  1-e^{-\int\nolimits_{t_0}^{t_1} \gamma_1^{\text{in}}(t) \, dt } \right) \right ]  e^{-\int_{t_1}^{t^{\ast}} \gamma_1^{\text{back}}(t) \,  d t } \,
\end{align}
where $P_{k}^{L}$ is the probability of the electron to be initially loaded into state $k$ by the time $t_0$, with $k=0, 1 , \ldots N$ (up to now we have considered $N=1$). 
The sequence of events contributing to the second term in Eq.~\eqref{eq:eq1} is shown schematically in Fig.~\ref{cmp}(c).

The initial probability distribution $P_{k}^{L}$ depends on the dynamics  prior to the onset of back-tunnelling.
Here we assume that the main QD is formed anew in each pumping cycle, and the chemical potentials of the parasitic states dwell below $\epsilon_0(t)$ once the QD is formed and occupied, as indicated schematically in Fig.~\ref{cmp}(c). Alternatively, one may consider a persistent QD and derive $P_{k}^{L}$ from the loading competition between the localized states as they get submerged under the Fermi level. We have modelled this alternative scenario and checked that it does not qualitatively affect the following analysis (see Supporting Information).
From here on, we focus on the simplest case of the electron always localized  in the intended QD initially ($P_0^{L}=1$, $P_{k}^{L}=0$ for $k>0$).

Since both in-tunnelling and back-tunnelling rates are controlled by the same tunnel barrier, the ratio
$\eta = \int_{t_0}^{t_1} \gamma_1^{\text{in}}(t) \, dt/ \int_{t_1}^{t^{\ast}} \gamma_1^{\text{back}}(t) \, dt $ can be approximated as independent of $v$. Although this is not strictly true (e.g. $\Delta t$ may depend on 
$v$ if the lever arm factors with respect to $V_\textup{C2}$ for state $0$ and state $1$ are different), this assumption does not change qualitatively the conclusions.
Setting $X_N=e^{-v}$ in Eq.~\eqref{eq:eq1} provides the following approximation for the pumped current due to loading and capture into a parasitic state: 
\begin{align} \label{eq:feature1}
 I_\textup{P}/(\mathrm{e}f) & = \exp [-e^{-v}] - \exp [-(1+\eta) e^{-v} ] \, .
\end{align}
The first term describes an increase in current due to reduced back-tunnelling while the second term describes a decrease in current due to reduced loading rates, as the gate voltage $v$ is tuned towards more positive values.
The maximum value of the analytic approximation in Eq.~\eqref{eq:feature1} with respect to $v$ is
\begin{align} \label{eq:Ipeak2}
   I_\textup{P}^{\text{peak}}/(\mathrm{e}f) = \eta \, (1+\eta)^{-1-\eta^{-1}} = \eta/e  + O(\eta^2) \, .
\end{align}
For a small $\Delta t$, we may approximate  
$\eta \approx [\Delta t \, \gamma_1^{\text{in}}(t_0)]/[\tau_1 \, \gamma_1^{\text{back}}(t_0)]$
where $\tau_1$ is 
the characteristic decoupling time for the parasitic state, $\gamma_1^{\text{in}}(t) \sim \gamma_1^{\text{back}}(t) \sim \exp (-t/\tau_1)$.
Hence,
$\eta$, while small, is directly proportional to the time delay between the dot levels crossing the Fermi energy.

A more complete model may take into account the continuous decay of occupation for states above the Fermi level concurrent with loading of the energetically lower parasitic states. The corresponding rate equations for $N$ parasitic states are rather straightforward to express, yielding an iterative solution in terms of $\gamma_k^{\text{back}}(t)$, $\gamma_k^{\text{in}}(t)$ and $t_k$, which involves nested integrals of increasing depth up to $N+2$ (see Supporting Information). These quadratures can be computed analytically if all $\gamma$ factors share the same time dependence. To illustrate the interplay of loading and capture, 
we solve the model with  $\gamma_k^{\text{back}}(t) =\tau^{-1} X_k \exp[-(t-t_k)/\tau]$ for $t>t_k$, and 
$\gamma_k^{\text{in}}(t) =g_k \gamma_k^{\text{back}}(t)$ for $t<t_k$, where $g_k$ is the quantum degeneracy factor of state $k$. We parametrize the rate and energy difference between the dots as $X_{k+1}/X_{k} = e^{-\delta}$ and 
$t_k = t_0 +k \, \Delta t$, respectively, and set $g_k=g=4$ (to account for spin and valley degeneracy).
With these parameters $\eta=(e^{\Delta t/\tau}-1) g$. The resulting expression for the current as a function of $v$ contains only
$\delta$, $\eta$, and $g$. Consequently, $N$ peaks are resolved for $\delta \geq 2.0$, with maxima following approximately Eq.~\eqref{eq:Ipeak2}.
We plot examples in Fig.~\ref{cmp}(b), with $N$, $\delta$, and $\eta$ chosen to yield similar features as those of the experimental data in panel (a). We find a good qualitative agreement between this simple theory and the experiments. \\\indent
For traces with a single maximum, we have used Eq.~\eqref{eq:Ipeak2} to estimate $\eta$ as a function of frequency, as shown in the inset of Fig.~\ref{cmp} (b). We find that $\eta(f)$ follows approximately a linear relation, once the first maximum develops at $f=f_c \approx 25\,$MHz.
Since our scenario for non-monotonic behaviour requires the state with a stronger confinement to be lower in energy, and $\eta \propto \Delta t$ once $\Delta t>0$, 
the observations suggest that the sequence in which the lowest energy levels of the competing states cross the Fermi level depends on frequency. A possible mechanism for this would be a small frequency-dependence of the lever arm factors of the parasitic states.\\\indent
Let us examine the high-speed operation of our multi-state pump and evaluate the precision of the generated current. Figure~\ref{accu}(a) shows the pumped current as a function of $V_\textup{C2}$ at $f=2.5$~GHz, corresponding to an absolute current at the plateau of roughly 400 pA. As discussed above, we tune the pump within the regime of complete emission and loading such that the capture error dominates. This is commonly regarded as the best operation point in terms of transfer accuracy within the single-QD decay cascade model.\cite{slava-rev} As shown in the inset of Fig.~\ref{accu}(a), at this frequency the distinct signature of the multistate-mediated initialisation is present. We are interested in determining whether the parasitic states exhibit detrimental effects on precision beyond the loading and capture model. The data shown in Fig.~\ref{accu}(b) have been acquired with an averaging procedure that records the current in cycles with the ac drive turned alternatively on and off.\cite{spg-ncomm} This allows one to remove instrument artefacts such as offset currents as well as  low-frequency noise and slow drifts during the relatively long averaging time. Although this technique is often used in conjunction with a null-detection set-up that compares the pumped current to a reference current traceable to primary standards,~\cite{spg-ncomm,mynano,gento-apl,gento-sr,zhao} we point out that in this experiment we did not use such a metrologically accurate measurement apparatus. Hence, our aim is merely to assess the flatness of the plateau at ppm level, as is possible with the small random uncertainty achieved by the employed averaging protocol. A comprehensive estimate of the accuracy of the pump would require a careful analysis of all the contributions to the systematic uncertainty of the measurement set-up, and lies beyond the scope of this work. Each data point in Fig.~\ref{accu}(b) has been acquired through 12 on/off cycles of 24 seconds each. The optimal duration of the cycles has been estimated upon verification that further improvements of the statistical uncertainty were not attained for longer timescales. A single half-cycle includes 120 points of which the initial 30 were systematically discarded to remove possible artefacts introduced by switching transients. The averaged current is evaluated as $I_\textup{HA}=\overline{I}_\textup{P}-\overline{I}_\textup{off}$, where $\overline{I}_\textup{P}$ is the mean of the ammeter readings during the ``on'' half-cycles and $\overline{I}_\textup{off}$ is the mean of the readings for the ``off'' half-cycles. The error bars represent the $1\sigma$ relative random uncertainty of each data set and are evaluated as the error of the mean of the normalized current, i.e. the standard deviation of 1080 points divided by $\sqrt{1081}$. The current plateau is identified by fitting the data with the single-state decay cascade model,~\cite{slava-prl,gento-sr} which provides a smooth extrapolation of the systematic $0$- and $2$-electron transfer error rates into the plateau region.
From the fit, one derives the point of inflection at ($-0.381$~V, $1.000438$). The plateau extension is then limited by the first point in each direction that is inconsistent with the value of the current at the inflection point using a 2$\sigma$ random uncertainty. The points that satisfy this criterion~\cite{stein-apl} are highlighted in yellow in Fig.~\ref{accu}(b). The 1$\sigma$ random uncertainty of these seven averaged data points is $0.46$~ppm and is an indication of the flatness of the plateau. 
Note that the deviation of the plateau-averaged current from the expected quantized value falls within the relative systematic uncertainty of the pre-amplifier gain calibration, which is quoted by the manufacturer as $10^{-3}$ for the conditions used in the experiments.

The origin of the additional states in competition with the intended QD during the initialization phase is likely to be ascribed to parasitic dots. 
Although the regular Coulomb diamond pattern of Fig.~\ref{set-up}(b) would suggest that a single QD determines the transport characteristics, one has to note that this is attained with a fairly large voltage applied to the entrance barrier, i.e. $V_\textup{IB}=0.78$~V. By contrast, for the pumping experiments, one has to tune the device into a different working point, namely $V_\textup{IB}$ needs to be significantly reduced to achieve robust quantization, i.e. $V_\textup{IB}<=0.51$~V. Consequently, one may unintentionally give rise to confinement potentials at the entrance barrier, leading to the formation of parasitic dots. These additional states may be undetected in dc transport because the entrance barrier is opaque, and the relevant tunnelling rate is too low to generate a measurable dc current. This type of parasitic states have been previously utilized to operate double-dot systems in planar MOS devices,\cite{menno,hyst} and can be attributed to the interplay between the electrostatic bias and strain-induced potentials.\cite{thor} Furthermore, the compact device design with a reduced number of gates used in this study may have played a role in the occurrence of the parasitic states. Indeed, by having a single top gate to control both the density of states in the leads and the dot potential, one has to set it at a relatively high voltage to ensure the continuity of the electron channels to the ohmics. This was compensated by lowering $V_\textup{IB}$ more than in previous studies~\cite{mynano,tuomo,scirep} in order to adjust the electrostatic landscape and to reach the few-electron regime in the pumping mode. This may have contributed to the formation of the parasitic states in the proximity of the entrance gate.\\\indent
Our data indicate that the effect of these states on the pumped current becomes more prominent as the pumping 	frequency increases.
In particular, we observe the onset of current non-monotonicity at $f_\textup{c} =25$~MHz and the linear growth of the peak height with $(f-f_\textup{c})$,  see the inset of Fig.~\ref{cmp}(b). We attribute this effect to 
the linear change in the parasitic state loading time $\Delta t(f) \propto (f-f_\textup{c})$, once $\epsilon_1(t_0)$ becomes less than $\epsilon_0(t_0)$ (and hence $\Delta t >0$) at $f=f_c$. Such situation arises naturally if capacitive coupling coefficients of the parasitic states depend on frequency: since our pumping scheme requires a large amplitude modulation, even a small change in the lever arm factors may affect the exact sequence in which the competing states emerge from the Fermi sea close to the minimum of the time-dependent potential (see Fig.~\ref{cmp}(c)). 
Strong temporal variation of the electron density in the vicinity of the modulated gate, and the resulting deviations from adiabaticity in screening of the disorder potential in the depleted QD region may be behind the observed frequency dependence. However, verification of such hypothesis is beyond the scope of the present study.
Unfortunately, our measurements do not provide direct quantitative information on the parameters used in the theoretical model. This is because our experiments do not probe the loading and ejection dynamics separately, owing to the simple sine wave used as a drive. In the future, we may extract more quantitative information by combining our model to experiments carried out with arbitrarily shaped waveforms~\cite{spg-ncomm,gento-ncomm} and precise control of rising and falling times at a constant driving frequency.
Nevertheless, these findings suggest that the interplay between disorder and screening may play a role in the understanding of high frequency pumping, a phenomenon not considered thus far in electron pumps.\\\indent Despite the complexity of the device used, we find it encouraging that the high-averaging measurements revealed a flat plateau within sub-ppm random uncertainty for gigahertz pumping. Although we cannot make an accurate claim on the absolute agreement between the generated current and the expected quantized value, our study reinforces the idea that pumps with unintended features, such as parasitic dots or trap states, may be of use for quantum metrology, whilst relaxing some of the demanding experimental requirements. Finally, in the wider context of quantum information, silicon systems hold high promise due to the achievement of uniquely long coherence times within the framework of an industrially viable technology. A deeper understanding of the interplay between engineered and parasitic localized states may enable scaling up small demonstrations of quantum control to large integrated systems.
\section*{Acknowledgement}
The authors thank R. Zhao, J. van der Heijden and L. Fricke for useful discussions. AR acknowledges support from the European Union's Horizon 2020 research and innovation programme under the Marie Sk\l{}odowska-Curie grant agreement No 654712 (SINHOPSI).  This work was also financially supported by the Australian National Fabrication Facility for device fabrication. JK and VK acknowledge support from University of Latvia grant no.\ AAP2016/B031. ASD acknowledges support from the Australian Research Council (DP160104923 and CE11E0001017), the US Army Research Office (W911NF-13-1-0024) and the Commonwealth Bank of Australia. MM acknowledges funding from the Academy of Finland through its Centres of Excellence Program (project no. 312300).  GCT acknowledges financial support from the ARC-Discovery Early Career Scheme Award (Single Atom Based Quantum Metrology - DE120100702) for the development of the set-up used in these experiments. This work has received funding from the European Metrology Programme for Innovation and Research (EMPIR, 15SIB08 e-SI-Amp) co-financed by the Participating States and from the European Union's Horizon 2020 research and innovation programme.

\pagebreak
\widetext
\begin{center}
\textbf{\large SUPPORTING INFORMATION\\[2in]Gigahertz Single-Electron Pumping Mediated by Parasitic States}
\end{center}
\setcounter{equation}{0}
\setcounter{figure}{0}
\setcounter{table}{0}
\setcounter{section}{0}
\setcounter{page}{1}
\makeatletter
\renewcommand{\theequation}{S\arabic{equation}}
\renewcommand{\thefigure}{S\arabic{figure}}
\renewcommand{\bibnumfmt}[1]{[S#1]}
\renewcommand{\citenumfont}[1]{S#1}
\renewcommand{\thesection}{S\Roman{section}}



\newpage
\thispagestyle{plain}
\section{Pump tunability at high frequency}

\begin{figure*}[h]
\includegraphics[scale=0.75]{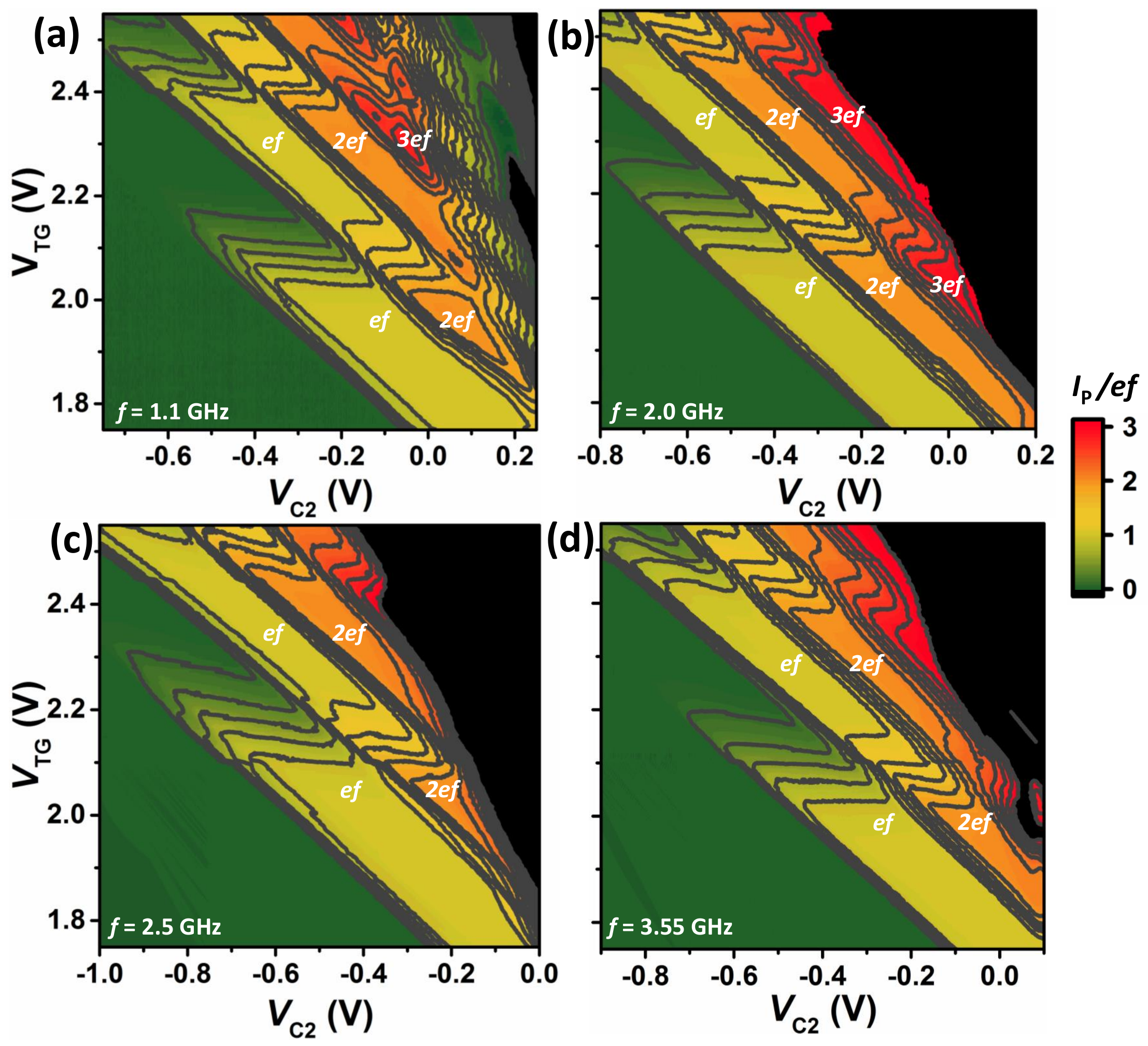}
\caption{Pumped current as a function of top gate and $\textup{C2}$ gate voltages. Experimental parameters: (a) $f=1.1$~GHz, $V_\textup{C1}=0.3$~V, $V_\textup{OB}=0.98$~V, $V_\textup{IB}=0.51$~V, and $P_\textup{ac}=3.6$~dBm. (b) $f=2.0$~GHz, $V_\textup{C1}=0.3$~V, $V_\textup{OB}=1.00$~V, $V_\textup{IB}=0.52$~V, and $P_\textup{ac}=3.5$~dBm. (c) $f=2.5$~GHz, $V_\textup{C1}=0.05$~V, $V_\textup{OB}=1.07$~V, $V_\textup{IB}=0.52$~V, and $P_\textup{ac}=3.7$~dBm. (d) $f=3.55$~GHz, $V_\textup{C1}=0.20$~V, $V_\textup{OB}=1.06$~V, $V_\textup{IB}=0.57$~V, and $P_\textup{ac}=5.0$~dBm.}
\label{fgr:checkers}
\end{figure*}

As discussed in the main article, it is important to operate the pump at a working point that ensures robust current quantisation. Tunable-barrier semiconductor pumps perform best when tuned at the first plateau with complete charge emission.~\cite{S-slava-rev} Figure~\ref{fgr:checkers} shows a significantly wider extension of the $1~ef$ plateau in these operating conditions, which confirms an enhanced robustness of the pumping protocol. Figure~\ref{fgr:checkers} also reveals that both the emission and capture limits can be tuned at gigahertz frequencies. However, at different frequencies, the optimal operation point may slightly shift, as one can observe by comparing the positions of the quantised current plateaus in the panels of Fig.~\ref{fgr:checkers}. This can be attributed to the frequency-dependent attenuations and reflections of the driving signal in the coaxial line of the cryostat. These non-idealities become usually more prominent for increasing frequency. As a result, in order to operate the pump, one has to increase the overall ac power at the signal generator and, consequently, re-adjust the dc voltages at some gate electrodes. Therefore, experiments similar to those of Fig.~\ref{fgr:checkers} have proved instrumental for coarse tuning the pump at gigahertz, in preparation for the time-consuming high-averaging measurements.
\thispagestyle{plain}

\section{Frequency dependence of the capture characteristics}

\begin{figure*}[t]
\includegraphics[scale=0.56]{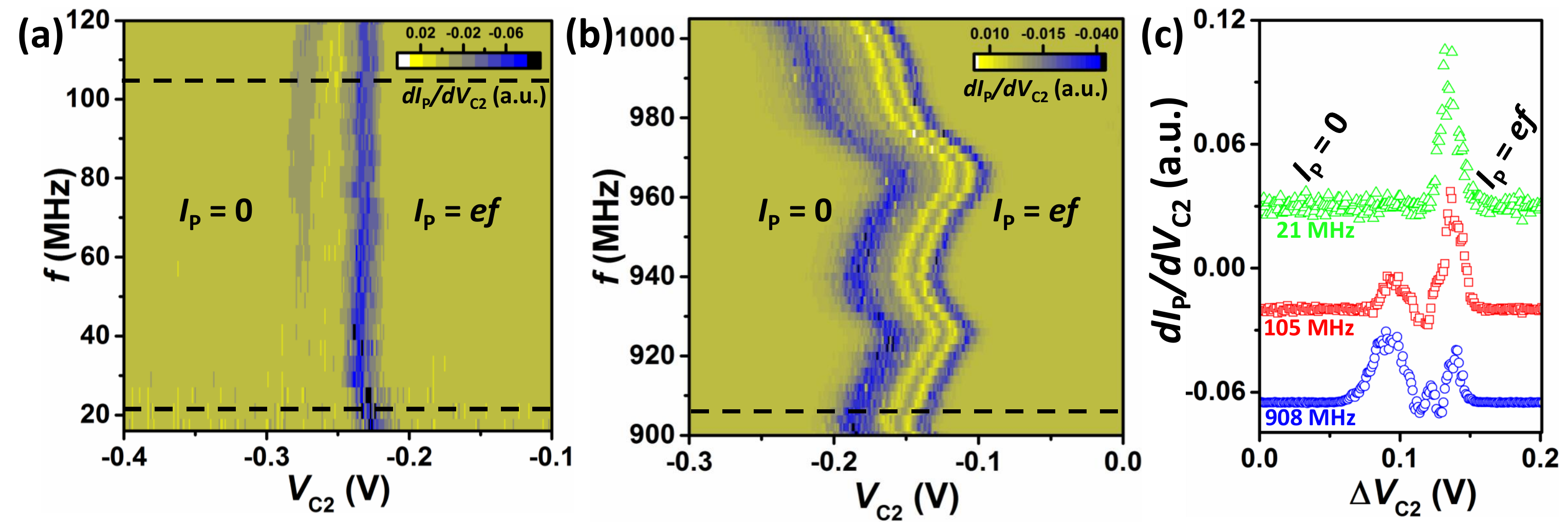}
\caption{Derivative of the pumped current with respect to gate $\textup{C2}$ voltage as a function of the driving frequency and gate $\textup{C2}$ voltage. Experimental parameters: (a) $V_\textup{TG}=2.0$~V, $V_\textup{C1}=0.1$~V, $V_\textup{OB}=1.2$~V, $V_\textup{IB}=0.52$~V, and $P_\textup{ac}=0.3$~dBm. (b) $V_\textup{TG}=1.9$~V, $V_\textup{C1}=0.2$~V, $V_\textup{OB}=0.98$~V, $V_\textup{IB}=0.51$~V, and $P_\textup{ac}=3.6$~dBm. (c) Traces denoted by dashed lines in panels (a) and (b) as functions of incremental voltage. The traces are shifted vertically for clarity by +0.03 for $f=21$~MHz, -0.02 for $f=105$~MHz and -0.07 for $f=908$~MHz.}
\label{fgr:freq}
\end{figure*}

As we discuss in the main article, the electron capture mechanisms are reflected in the rising edge of the current plateau, which shows a distinctive frequency dependence in the experiments discussed. As shown in Fig.~\ref{fgr:freq}(a), at low frequency the current rise is monotonic, which results in a single peak in the first derivative of the current with respect to $V_\textup{C2}$. However, a second peak emerges for increasing frequency, and a third one appears at furtherly high frequency, as illustrated in Fig.~\ref{fgr:freq}(b). Note that the previously discussed effects arising from reflection patterns in the coaxial cables are clearly visible by comparing panels (a) and (b). Indeed, the capture boundary line in Fig.~\ref{fgr:freq}(a) appears essentially at the same gate voltage for all frequencies, but Fig.~\ref{fgr:freq}(b) reveals significant horizontal dependence for a comparable frequency range.\\\indent
In the previous literature,~\cite{S-kata,S-mynano} peaks in the derivative of the pumped current were associated to excited QD states through which the electron escape was enhanced. This interpretation does not satisfactorily explain the findings presented here. Figure~\ref{fgr:freq}(c) shows the first derivatives of the traces reported in Fig. 3(a) of the main text. As discussed in the main text, we argue that the extra peaks that appear for increasing frequency are the signature of states that compete for capturing the electron, as opposed to merely contributing to an enhancement of the escape rates.

\section{Model}

\subsection{Definition of the capture and loading probabilities}

Here we consider a general model for the pumped current which includes the results reported in the main text as special cases. The full initialization phase consists of loading (L), followed by capture (C).
We enumerate states in the order with which they emerge from the Fermi sea during capture, starting with the main dot $k=0$, following by $k=1\, \ldots N$ parasitic states. The total current in the capture-limited regime is
  \begin{align} \label{eq:currentGeneral}
   I_\textup{P}/(\mathrm{e}f) = \sum_{k=0}^{N}  P^{\text{C}}_k \, ,
  \end{align}
where
$P_k^{\text{C}}$ is the probability for 
the electron to be captured and eventually delivered to the drain in state $k$.
We compute $P_k^{\text{C}}=P_k(t^{\ast})$ as  the final value of the time-dependent occupation probability at time $t^{\ast}$ when the entrance barrier is so high that all charge exchange with the source effectively ceases.
The initial condition  $P_k(t_0)=P_j^{\text{L}}$ is the probability for the state $k$ to be loaded before any back-tunnelling starts at $t_0$.

We first solve the capture problem for arbitrary $P_k^{\text{L}}$ subject only to 
complete initial loading condition $\sum_{k=0}^{N} P_k^{\text{L}}=1$. Next, we solve separately the problem of initial loading, assuming that the time-dependence of the only parameter controlling both energies and rates is symmetric with respect to $t=0$. This condition holds for harmonic modulation of the entrance barrier voltage $V_{\text{IB}}(t)$ with $t=0$ corresponding to the most positive value during the pumping cycle, which separates the loading phase at $t<0$ from the capture phase at $t>0$.

We characterise the tunnel coupling between the source and a state $k$ by the in-tunnelling ($\gamma_k^{\text{in}}$) and the back-tunnelling ($\gamma_k^{\text{back}}$) rates, defined by the Fermi golden rule without the Pauli blocking factors as
$\gamma_k^{\text{in}} =g_k \gamma_k^{\text{back}} =g_k 2 \pi \langle \rho |M |^2 \rangle_k $, where $g_k$ is the quantum degeneracy of the discrete state $k$, $\rho$ is the density of continuous states in the lead, and $M$ is the tunnelling matrix element;  the averaging is done over the lead conduction modes at energy $\epsilon_k$.

\subsection{Solution for the capture problem}
We denote $t_k$ the time moment when the energy of a state $k$ crosses the Fermi level on the way up.
Consider a particular time interval $t \in [t_j \ldots t_{j+1}]$ for $j=0 \ldots N-1$. 
The rate equation for the probability $P_k(t)$ of the state $k$ to be occupied
is dominated by back-tunnelling of electrons for $k \le j$ and by in-tunnelling (loading) for $k>j$,
\begin{align} \label{eq:C1}
 0\le k \le j: & & \frac{d}{dt}P_k(t) & =  -\gamma_k^{\text{back}}(t) \, P_k(t) \,  , \\
 j < k \le N: & & \frac{d}{dt}P_k(t) & =  +\gamma_k^{\text{in}}(t) \, \left [1 - \sum_{k=0}^{N} P_k(t)  \right]  \, . 
 \label{eq:C2} 
\end{align}
The initial condition is $P_k(t_0) =P^{L}_{k}$.
Equations~\eqref{eq:C1} and \eqref{eq:C2} disregard the effect of thermal fluctuations in the lead, and make Markovian assumptions in the use of time-dependent rates.   

Time evolution from $t=t_j$ to $t_{j+1}$ defined by  Eqs.~\eqref{eq:C1} and \eqref{eq:C2} can be expressed explicitly as
\begin{align} \label{eq:PCfirst}
 0 \le k \le j: & &  \quad P_k(t_{j+1}) & = P_{k}(t_k) \exp \left [ - \int_{t_k}^{t_{j+1}} \gamma_k^{\text{back}}(t') \, d t \right ] \\
 j < k \le N: & &  \quad P_k(t_{j+1}) & = P_k(t_{j}) + \int_{t_j}^{t_{j+1}} \gamma_k^{\text{in}}(t)  
\Bigl \{ [1-\sum_{l=j+1}^{N} P_l(t_j)  ]  
\exp \left [-\int_{t_j}^{t} W_j (t') dt'\right ] \, ,\label{eq:complex} \\
 &&  &  + 
   \int_{t_j}^{t} S_j(t') W_j(t')\exp \left [-\int_{t'}^{t} W_j (t'') dt''\right ] d t' - S_j(t) \Bigr \} \, d t  \, , \nonumber 
\end{align}
where
\begin{align}
  W_j(t) & = \sum_{k=j+1}^{N} \gamma_k^{\text{in}}(t) \, ,\\
  S_j(t) & = \sum_{k=0}^j P_{k}(t_k)  \exp \left [ - \int_{t_k}^{t} \gamma_k^{\text{back}}(t') \, d t \right ] \, .
\end{align}

For $t>t_N$, only back-tunnelling remains possible for any of the states,
\begin{align} \label{eq:PClast}
  P_k^{C}=P_{k}(t^{\ast}) =P_{k}(t_N)  \exp \left [ - \int_{t_N}^{t^{\ast}} \gamma_k^{\text{back}}(t') \, d t \right ] \, .
\end{align}

For rates that depend exponentially on time, as defined in the main text, Eq.~\eqref{eq:complex} can be integrated leading to finite but cumbersome algebraic expressions in terms of $P_{k}^{L}$, $\delta$, $\Delta t/\tau$ and $\eta$.

For $N=1$, the explicit form is
\begin{align}
   P_0^{C} = &  \exp{(-e^{-v+\delta})} P_0^{L} \, , \\
 P_1^{C} =  & \exp{(-e^{-v})} \left ( 
 1 -   \exp{[-\eta \, e^{-v}]} P_0^{L} + \eta \, e^{-\delta+\Delta t/\tau}
\frac{ \exp\{-e^{-v+\delta}(1 -e^{-\Delta t/\tau})] \}-\exp{[-\eta \, e^{-v}]} }{e^{\Delta t/\tau} -1-\eta \, e^{-\delta+\Delta t/\tau}} \, P_0^{L} \right ) 
\, .
 \label{eq:PCN1explicit} 
\end{align}
In the large $\delta$ limit, the above equations simplify to $P_0^{C}=0$ and $P_1^{C}=\exp{(-e^{-v})} -\exp{[-(1+\eta) \, e^{-v}]} P_0^{L}$.
Taking into account that $Z_1= \eta X_1=\eta \,  e^{-v}$, and $P_{0}^{L}+P_{1}^{L}=1$, and using Eq.~\eqref{eq:currentGeneral}, 
we see that in this limit the result agrees with Eq.~(1) of the main text.

\begin{figure*}[htb]
\includegraphics[width=0.75\textwidth]{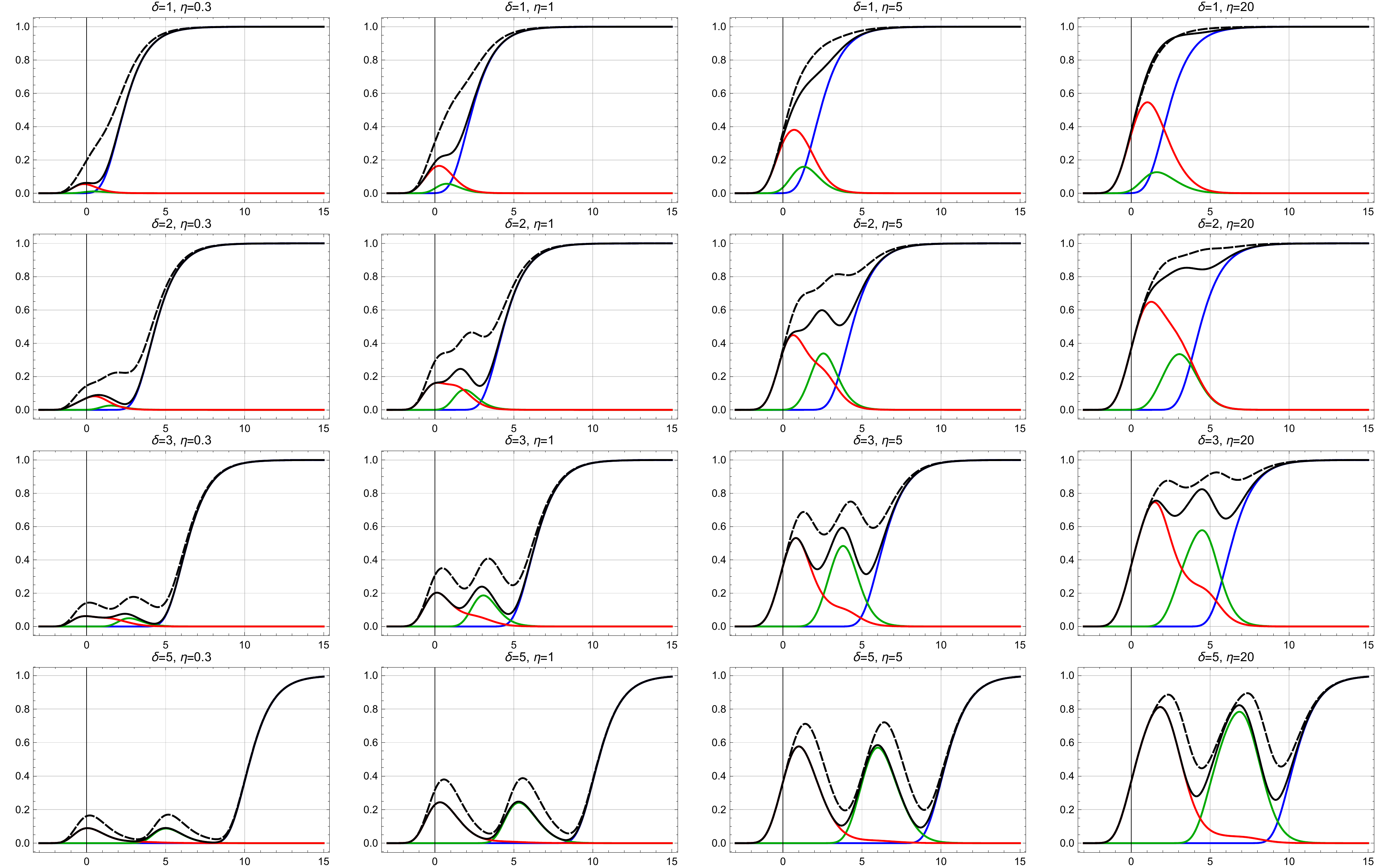}
\caption{Modelled pumped current $I_{P}/\mathrm{e} f$ as a function of dimensionless gate voltage $v$ for $N=2$, $g=4$ and different combinations of $\delta$ and $\eta$. The continuous black line shows the total current, computed from Eq.~\eqref{eq:currentGeneral} using the solution \eqref{eq:PCfirst}-\eqref{eq:PClast} to the capture problem, and assuming $P_{0}^{L}=1$, $P_{k>0}^{L}=0$.
Red, green and blue lines show components corresponding to $P_{2}^{C}$, $P_{1}^{C}$ and $P_{0}^{C}$, respectively.
The dashed black line shows the total current, computed using the same expressions for $P_k^{C}$, but for a different distribution of the initial loading probabilities, $P_{k}^{L}$, as given by Eqs.~\eqref{eq:PLspecial}.}
\label{fgr:model2D}
\end{figure*}

In Fig.~\ref{fgr:model2D}, we show characteristic behaviour of the total current for $N=2$, $g=4$ and a range of values for $\delta$ and $\eta$.
In general, the parameter $\delta$ controls the distance between features corresponding to the different states, and $\eta$ regulates the height of the local maximum.

\thispagestyle{plain}

\subsection{Solution for the loading problem}

The loading starts at $t=t^{L}_N<0$ when all the states are empty and the state with the lowest time-dependent energy $\epsilon_N(t)$
crosses the source Fermi level, $\epsilon_N(t^{L}_N)=0$, on the way down, $\dot{\epsilon}_N(t^{L}_N) <0$. This is followed 
by other states at $t^{L}_{N-1} < t^{L}_{N-2} < \ldots t^{L}_{0}$ such that $\epsilon_k(t^{L}_k)=0$ and $t^{L}_0 <0$.

The rate equation for the probability $P_k^{\text{L}}(t)$ of the state $k$ to be loaded by the time $t$ 
in the low-temperature limit can be written as 
\begin{align} \label{eq:kinetic}
\frac{d}{dt} P_k^{\text{L}}(t) &=  \Theta(t-t^{L}_k) \gamma_{k}^{\text{in}}(t) 
\left [ 1- \sum_{l=0}^{N} P_l^{\text{L}}(t) \right ]  -  \Theta(t^{L}_k-t) \gamma^{\text{back}}_{k}(t)  \,  P_k^{\text{L}}(t) ,
\end{align}
where  $\Theta(t-t^{L}_j)$ is the Heaviside step function. The initial condition $P_k(t)=0$ for $t<t^{L}_k$ reflects the assumption 
of complete emission in the previous pumping cycle.

Summing up Eqs.~\eqref{eq:kinetic} for all $k$ gives a straightforward solution for the total instantaneous loading probability, $\sum_{l=0}^{N}  P_l^{\text{L}}(t)=1- \exp\left [ - Y(t) \right]$, 
in terms of $Y(t)= \sum_{j=0}^{N} \Theta(t-t^{L}_j) \int^{t}_{t^{L}_j}  \gamma_{j}^{\text{in}}(t') d t'$. Consequently, the individual probabilities $P_k^{\text{L}}(t)$ are given by
\begin{align}
P^{\text{L}}_k(t) & = \Theta(t-t^{L}_k) \int_{t^{L}_k}^{t} \gamma_k^{\text{in}}(t') \exp\left [ - Y(t') \right] dt' \, .
\end{align}

A common time dependence for individual rates, $\gamma_{k}^{\text{in}}(t) / \gamma_{N}^{\text{in}}(t) = \text{const} =K_{k}-K_{k+1}$,
allows us to compute $Y(t)$ and $P^{\text{L}}_k(t)$ by piecewise integration over time intervals $t\in [t^{L}_{j} \ldots t^{L}_{j-1}]$ in which  
$Y(t) =Y(t^{L}_j) + K_j \int_{t^{L}_j}^t \gamma_N^{\text{in}}(t') \, d t'$. The result can be expressed in terms of
 $Y_j=Y(t^{L}_j)=\sum_{l=j+1}^{N} K_l Z_l$ and  $Z_l =\int_{t^{L}_l}^{t^{L}_{l-1}} \gamma_N^{\text{in}}(t) \, dt$,
\begin{align} \label{eq:Plgeneral}
  P^{\text{L}}_k = & (K_{k}-K_{k+1})\sum_{j=0}^{k}K_j^{-1} \left ( e^{-Y_j}- e^{-Y_{j-1}} \right) \, ,
\end{align}
where we set $Y_N=K_{N+1}= 0$ and $Y_{-1}=+\infty$ for notational consistency. 
Note that  $K_l=\sum_{j=l}^{N} \gamma_{j}^{\text{in}}/\gamma^{\text{in}}_{N}$.

Explicitly, for $N=1$ and $N=2$,
\begin{align}
  & (N =1) &  P_0^{L} = & (K_0-K_1) K_0^{-1}  e^{- K_1 Z_1} \, , &  P_1^{L}= 1- P_0^{L} \, , &  \label{eq:PL1} \\
  & (N =2) &  P_0^{L} = &  (K_0-K_1) K_0^{-1} e^{- K_1 Z_1-K_2 Z_2} \, , &     \nonumber \\
  &  & P_1^{L} = & [K_1-K_2] \left [ K_0^{-1} e^{- K_1 Z_1} + K_1^{-1} (1- e^{- K_1 Z_1} )\right ] e^{-K_2 Z_2}\, , & 
  P_2^{L}= 1- P_0^{L}-P_1^{L} \, . \label{eq:PL2} 
\end{align}

\subsection{Connection between the loading and the capture problem}



Due to time-reversal symmetry of the sinusoidal single-parametric driving, we expect $t^{L}_k=-t_k$,
$\epsilon_k(t)=\epsilon_k(-t)$,  and 
$\gamma_k^{\text{in}}(t)= g_k^{} \gamma^{\text{back}}_k(-t)$.

This connects the parameters of the loading  to those of the capture,
\begin{align} \label{eq:ZXconnection}
   Z_k =  ({\gamma_N^{\text{in}}}/{\gamma_{k-1}^{\text{in}}}) \, g_{k-1} X_{k-1} - ({\gamma_N^{\text{in}}}/{\gamma_{k}^{\text{in}}}) g_{k} \, X_k \, .
\end{align}

In terms of the parameters $\delta$, $\Delta/\tau$ and $\eta$ defined in the main text, 
Eqs.~\eqref{eq:PL1} and \eqref{eq:PL2} can be written explicitly as
\begin{align}
  & (N =1) &  P_0^{L} = & \frac{\exp( - \eta \, e^{-v})}{1+e^{-\delta+\Delta t/\tau}} \, , \label{eq:P0LN1} \\
  & (N =2) &  P_0^{L} = & \frac{\exp[- \eta \, e^{-v} (1+e^{\delta}+e^{\Delta t/\tau})]}
  {1-e^{-\delta+\Delta t/\tau}+e^{-2 \delta+2\Delta t/\tau}}  \, , \label{eq:PLspecial} \\
  &  &  P_1^{L}= & \frac{\exp(- \eta \, e^{-v} )}{1+ 
   e^{-\delta+\Delta t/\tau}}- \frac{\exp[- \eta \, e^{-v} (1+e^{\delta}+e^{\Delta t/\tau})]}
  {(1+ 
   e^{-\delta+\Delta t/\tau})(1+ 
   e^{-\delta+\Delta t/\tau}+e^{-2\delta+2 \Delta t/\tau})} \, .
   \nonumber 
\end{align}
\begin{figure*}[b]
\includegraphics[scale=0.35]{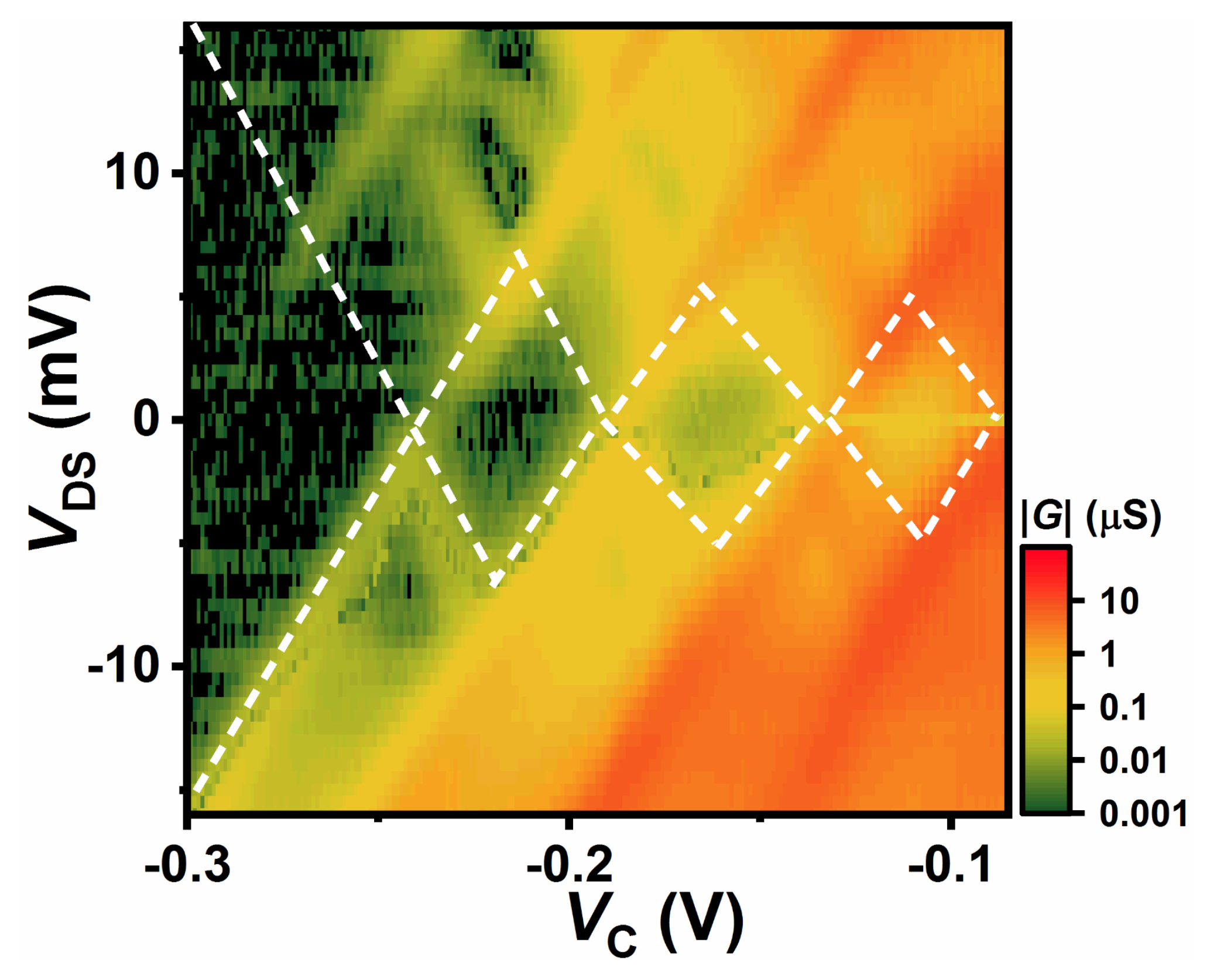}
\caption{Absolute differential conductance of the device as a function of source-drain and confinement gate voltages ($V_\textup{C}=V_\textup{C1}=V_\textup{C2}$). Dashed lines are guides for the eye to highlight the boundaries of the Coulomb diamonds. The other gate voltages are: $V_\textup{TG}=3.00$~V, $V_\textup{OB}=0.99$~V, $V_\textup{IB}=0.87$~V.}
\label{fgr:diam}
\end{figure*}
In general, competition between different states during the loading phase (as described by $P_{k}^{L}$) leads to results, similar to competition at the capture stage only (assumes $P_{0}^{L}=0$), see comparison in Fig.~\eqref{fgr:model2D}. For the simple case of $N=1$ and large $\delta$, large $X_0$ limit, 
Eq.~\eqref{eq:P0LN1} gives $P_0^{L} =\exp( - \eta \, e^{-v}) =e^{-Z_1}$. Using this value instead of $P_0^{L}=1$ in Eq.~(1) of the main text leads
to Eqs.~(2) and (3) with the value $\eta$ replaced by $2 \, \eta$, ie. for a single, well-pronounced parasitic state feature the difference between the two 
loading scenarios is just the doubling of the parameter $\eta$.
\section{Coulomb Blockade}
The compact device architecture that we have shown in Fig. 1(a) of the main manuscript can be operated as an error-resilient pump as long as a QD is formed in the region enclosed by the barrier and confinement gates. The desirably large charging energy observed in the Coulomb diamond plot as a function of $V_\textup{TG}$ (see Fig. 1(b) of the main manuscript) suggests that a very small QD is formed within the transport channel. However, since the top gate extends across the whole two-dimensional electron gas between the ohmics, one needs to perform additional transport measurements to verify that the QD is indeed formed in the region of interest. Figure~\ref{fgr:diam} shows a Coulomb diamond plot as a function of both confinement gates being kept at the same values and swept simultaneously. From this measurement and similar ones as a function of the barrier gate voltages, one can infer that an intended QD is indeed formed in the region highlighted in red in the inset of Fig. 1(a) of the main manuscript.

\end{document}